\documentclass[preprint]{aastex}
\usepackage{natbib}
\def\4u{4U1608-522}
\def\gx{4U1728-34}
\def\1636{4U1636-536}

\slugcomment{Submitted to the astrophysical journal}
\shorttitle{Spectral-timing of kHz QPOs}
\shortauthors{Peille et al.}

\begin{document}

\title{The spectral-timing properties of upper and lower kHz QPOs}

\author{Philippe Peille\altaffilmark{1,2},  Didier Barret\altaffilmark{1,2} and Phil Uttley\altaffilmark{3}}
\affil{Universit\'e de Toulouse; UPS-OMP; IRAP; Toulouse, France}
\affil{CNRS; Institut de Recherche en Astrophysique et Plan\'etologie; 9 Av. colonel Roche, BP 44346, F-31028 Toulouse cedex 4, France}
\affil{Anton Pannekoek Institute, University of Amsterdam, Postbus 94249, 1090 GE Amsterdam, The Netherlands}
\email{philippe.peille@irap.omp.eu}

\begin{abstract}
Soft lags from the emission of the lower kilohertz quasi-periodic oscillations (kHz QPOs) of neutron star low mass X-ray binaries have been reported from \4u\ and \1636. Those lags hold prospects for constraining the origin of the QPO emission. In this paper, we investigate the spectral-timing properties of both the lower and upper kHz QPOs from the neutron star binary \gx, using the entire \emph{Rossi} X-ray Timing Explorer archive on this source. We show that the lag-energy spectra of the two QPOs are systematically different: while the lower kHz QPO shows soft lags, the upper kHz QPO shows either a flat lag-energy spectrum or hard variations lagging softer variations. This suggests two different QPO-generation mechanisms.  We also performed the first spectral deconvolution of the covariance spectra of both kHz QPOs. The QPO spectra are consistent with Comptonized blackbody emission, similar to the one found in the time-averaged spectrum, but with a higher seed-photon temperature, suggesting that a more compact inner region of the Comptonization layer (boundary/spreading layer, corona) is responsible for the QPO emission. Considering our results together with other recent findings, this leads us to the hypothesis that the lower kHz QPO signal is generated by coherent oscillations of the compact boundary layer region itself. The upper kHz QPO signal may then be linked to less-coherent accretion-rate variations produced in the inner accretion disk, being detected when they reach the boundary layer.
\end{abstract}


\section{Introduction}

Since their first detection in neutron star low-mass X-ray binaries soon after the launch of the \emph{Rossi X-ray Timing Explorer} \citep[RXTE;][]{Bradt:1993,van-der-Klis:1996,Strohmayer:1996}, kilohertz quasi-periodic oscillations (kHz QPOs) have generated much interest. They are considered as a potential probe of General Relativity, due to the commensurability of their frequencies with the orbital frequency of matter very close to the neutron star surface. Until now, no model can account for all the properties of these oscillations \citep[see][for a review]{van-der-Klis:2006} and an application to testing strong field General Relativity remains speculative. Until recently, most studies on kHz QPOs focussed on their frequency, quality factor, spectral and rms amplitude properties. However, \cite{Barret:2013} and \cite{de-Avellar:2013} recently reintroduced energy dependent time lags long after the first detection of kHz QPO soft lags (variations in soft photons lagging those of hard photons) in the early years of RXTE \citep{Vaughan:1998}. In active galactic nuclei (AGN), high frequency soft lags are now routinely detected \citep{Tripathi:2011,de-Marco:2011,Emmanoulopoulos:2011,Zoghbi:2011,Cackett:2013}. Broad Fe K lags have also been reported from several AGN and interpreted as due to the light travel time between the central hard irradiating source and the reverberating disk (\citealt{Zoghbi:2012,Kara:2013a,Kara:2013} and see \citealt{Uttley:2014} for a review).  Soft lags were also seen at frequencies above 1~Hz in the black hole X-ray binary GX~339-4 and were attributed to reverberation of blackbody emission from the disk in response to driving Comptonized continuum variations \citep{Uttley:2011}.

Soft lags for the lower kHz QPO were found in \4u, as well as hints for reverberation from the detection of a bump around the Fe line energy in the highest quality lag-energy spectrum \citep{Barret:2013}. These soft lags were also reported in \1636\ by \cite{de-Avellar:2013}. Because the lower kHz QPOs are easier to detect, owing mostly to their higher quality factor compared to the upper kHz QPOs, very little is known about the lag properties of the latter. Nonetheless, \cite{de-Avellar:2013} reported a constant hard lag in the upper kHz QPOs from 4U1636-536 (but a lag consistent with zero from \4u). The RXTE data archives contain a wealth of observations with kHz QPOs from a large sample of neutron star systems. The detection of lags suggests that we should take a fresh look at these observations, using a wider range of spectral-timing techniques, because these data may enable us to understand where the kHz QPOs are produced. In this paper, we investigate the spectral-timing properties of the two kHz QPOs from \gx, for which the duty cycles of the lower and upper kHz QPO appearance in continuous data sets (ObsID in RXTE terminology) are comparable.

\section{Data analysis}
We have retrieved all the archival observations of \gx. In order to systematically detect the QPOs, the data in each Event File is segmented in intervals of 1024 seconds, and mean Fourier Power Density Spectra (PDS) are computed as the average of 1024 PDS integrated over 1 second (with 1/2$^{12}$ s time resolution, in the 3-30~keV band). X-ray bursts, as well as count rate drops, were carefully removed from the data. QPOs are searched for as excess powers between 550 and 1300 Hz and fitted with the Maximum Likelihood method of \cite{Barret:2012}. In each segment, zero, one, or two QPOs are thus detected and characterized. A QPO is considered significantly detected when the ratio between the Lorentzian normalization and its $1 \sigma$ error is larger than 3 \citep{Boutelier:2010}.

\subsection{QPO identification}
Among the properties of high-frequency QPOs, it is well known that the lower and upper kHz QPOs follow different tracks in a quality factor versus frequency diagram \citep{Barret:2006}. In these diagrams, the lower kHz QPO shows larger quality factors rising with frequency (and dropping rapidly at the highest frequencies), while the upper kHz QPO shows smaller quality factors which follow a smooth rising trend with frequency. In this paper, we identify kHz QPOs based on their position in the quality factor$-$frequency diagram.

\begin{figure}
\epsscale{1}\plotone{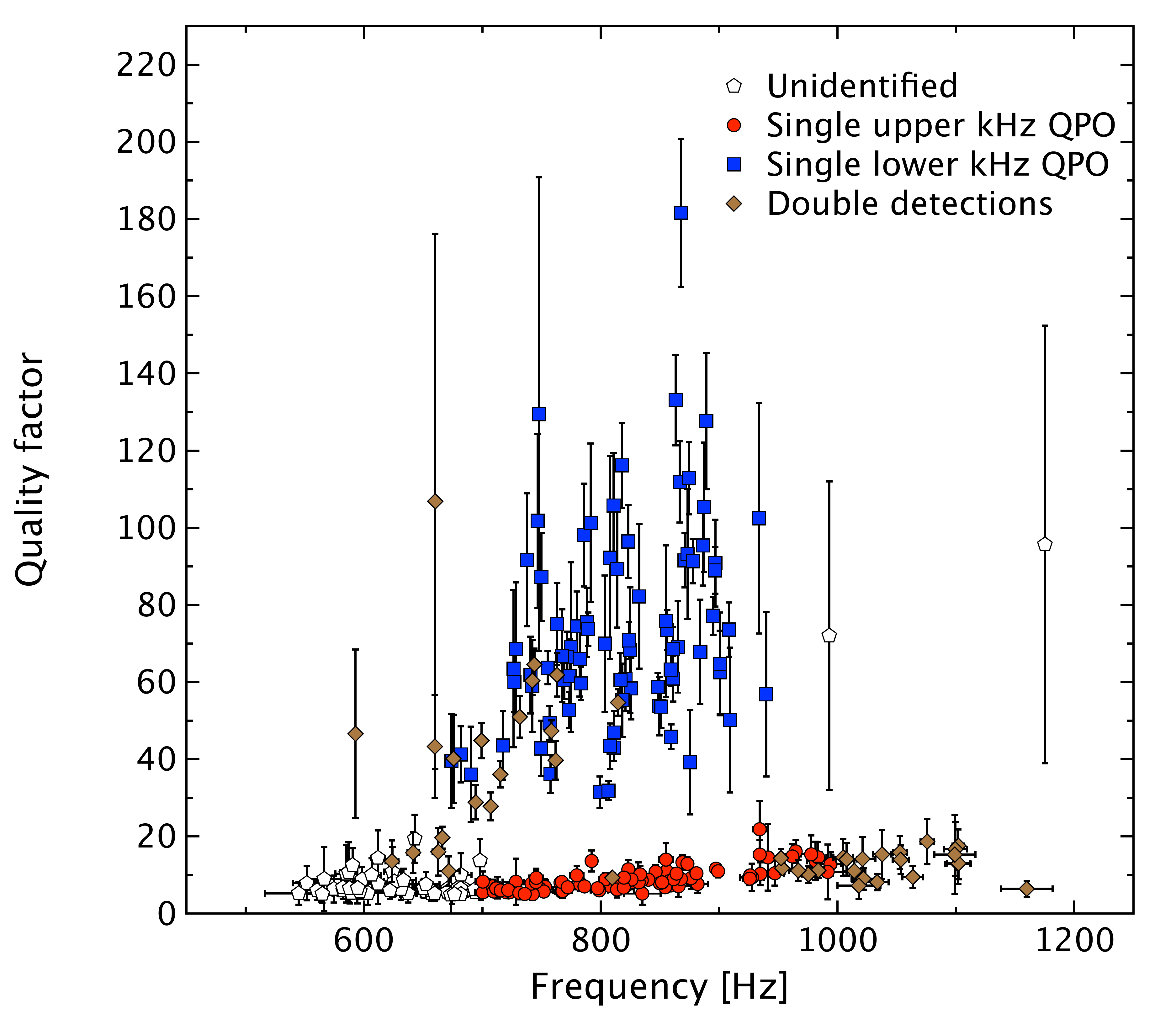}
\caption{Quality factors of all the kHz QPOs detected from \gx\ on timescale of 1024 seconds. Each point corresponds to one Event File. In this diagram, lower and upper kHz QPOs follow two different branches and can thus be identified from their position. QPOs falling in between the two groups of QPOs are not considered (listed as unidentified). \label{Q_vs_freq}}
\end{figure}

For this purpose, within each Event File, all the segments of data in which a QPO is significantly detected are shifted and added with respect to the mean QPO frequency.  In case two QPOs were detected, the one with the highest duty cycle is chosen, by default the one of lower frequency. The Event File averaged shifted-and-added PDS is then searched for excess powers, and one or two QPOs (the lower and/or the upper) are fitted and quality factor (Q) derived. This produces one or two values of Q per Event File. The quality factors of all the QPOs detected from \gx\ are shown in Figure \ref{Q_vs_freq}. As can be seen, the lower and upper kHz QPO branches are easy to distinguish, allowing us to identify all Event Files as containing either a lower, or an upper or even twin kHz QPOs. Taking a conservative approach, we leave out segments of data containing QPOs with quality factor less than 5 and QPOs at the boundaries between the two branches. For Event Files in which a single QPO is detected, we define lower kHz QPOs as those with frequencies between 650 and 950 Hz, and quality factors larger than 30. Similarly upper kHz QPOs are defined as those with frequencies between 700 and 1200 Hz and quality factor between 5 and 25. The identification of the lower and upper kHz QPO is obvious in the 24 Event Files (corresponding to 17 ObsIDs) in which they are simultaneously detected (maroon diamonds in Fig. \ref{Q_vs_freq}). Detections with quality factors below 25 at frequencies smaller than 700~Hz were ignored as they correspond to an overlap of the quality factor versus frequency correlations of both QPOs, hampering their identification in a reliable way.

\subsection{Frequency dependent lags}
\begin{figure}
\epsscale{0.65}
\plotone{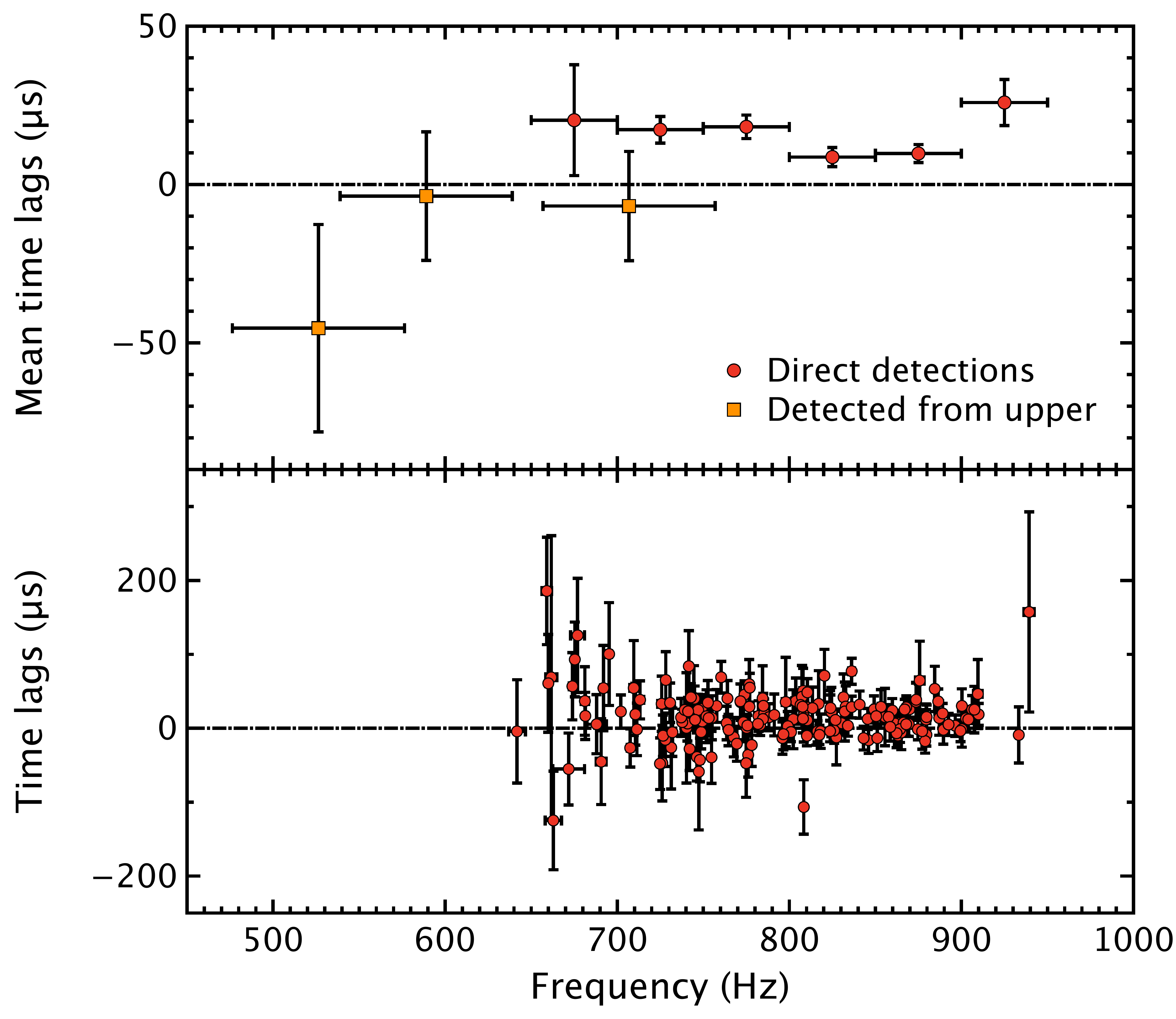}
\plotone{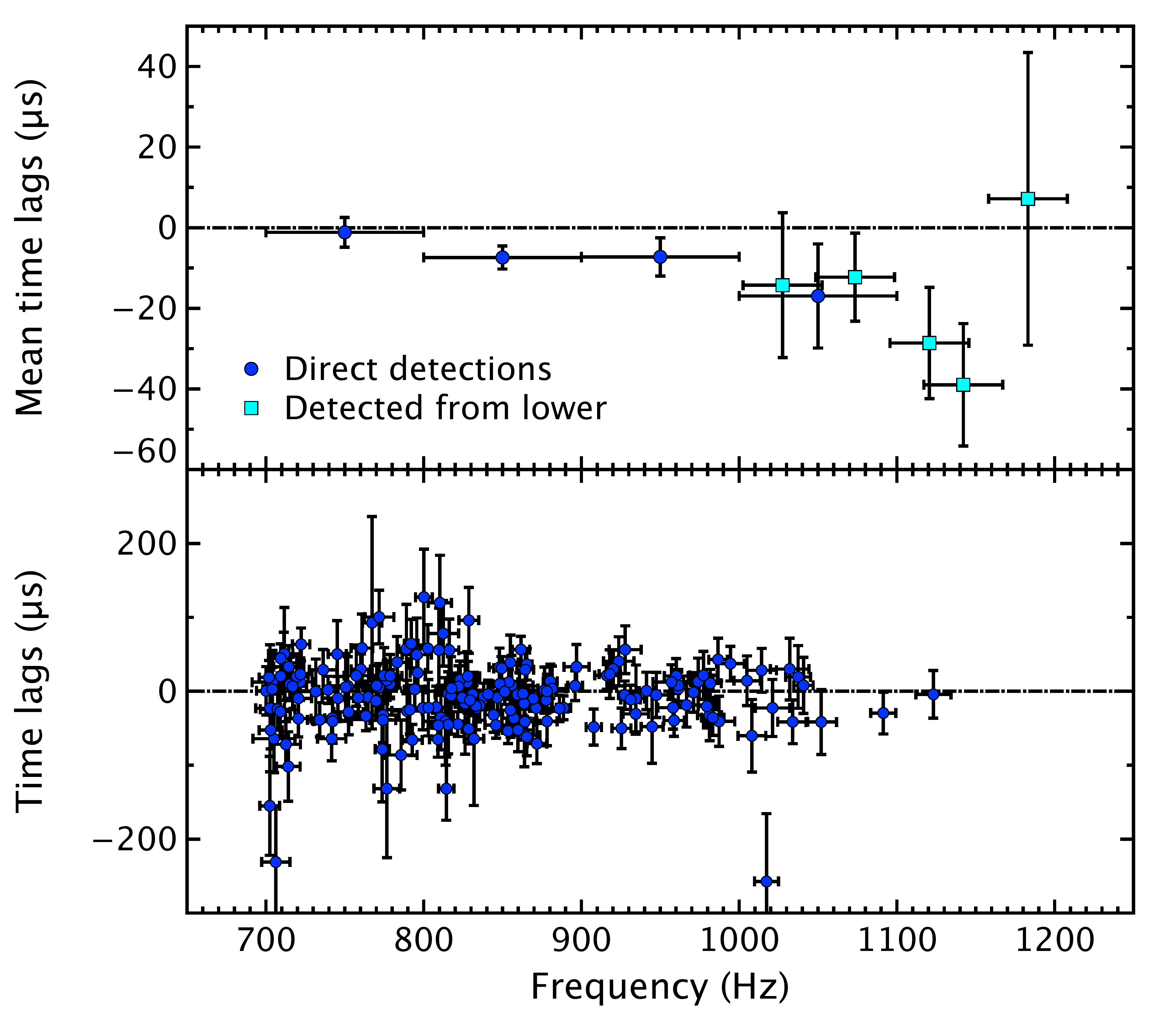}
\caption{Time lags between the 3$-$8~keV and 8$-$30~keV time series as a function of frequency for the lower (top) and upper (bottom) kHz QPO from \gx. A soft lag is positive. For each QPO, the upper part of the figure shows the distribution of the lags computed after shifting and adding all the detected QPOs inside adjacent frequency bins (circles). To expand the frequency range, the PDS containing an upper kHz QPO were shifted and added as to obtain a significant detection of the lower kHz QPO, and vice-versa (squares). The bottom parts show the lags computed from the original individual detections on 1024~s segments. Clearly the lag behavior of the two QPOs is different. While the lower kHz QPO shows a soft lag, a hard lag is suggested for the upper kHz QPO. \label{lag_vs_freq}}
\end{figure}

For computing the lags, we follow Barret (2013). Namely, we compute the lags between the 3--8 keV and 8--30 keV emission, over continuous data segments of each Event File. We average 1024 segments of 1 second duration and obtain the lags from the cross-spectrum, averaged over the frequency range covered by the full-width at half maximum (FWHM) of the QPO Lorentzian profile, as derived from fitting the QPO profile averaged over the 1024 segment PDS. To display trends otherwise not visible, the lags are also computed to higher signal-to-noise from the shifted-and-added cross spectra averaged over broader QPO frequency ranges. In order to correct for dead-time induced cross talk, we subtracted from each cross spectrum the average Fourier amplitude in the $1350-1700$~Hz band where Poisson fluctuations are the only source of variance, as suggested by \cite{van-der-Klis:1987}. The lags of both the lower and upper kHz QPOs are shown in Fig.~\ref{lag_vs_freq}, together with the frequency averaged values. From that figure, it is clear that the lower kHz QPO shows a soft lag, while for the upper kHz QPO, there is an indication of a hard lag at high frequency. The soft lags for the lower kHz QPOs of \gx , which are reported for the first time here, are comparable to those of \1636\ \citep{de-Avellar:2013} and about twice as small as the lags of \4u \citep{Barret:2013,de-Avellar:2013}.

One can also extend the frequency range for the lag measurements by shifting-and-adding the PDS containing a lower kHz QPO so as to obtain a significant detection of the upper kHz QPO. This covers the upper frequency range spanned by the upper kHz QPO (above 1000 Hz). The lags so obtained are shown in Fig. \ref{lag_vs_freq}, indicating also a hard lag for the upper kHz QPO, although the error bars are quite large due to the weakness of the QPO signal (the RMS amplitude of the upper kHz QPO keeps decreasing with frequency). A constant time lag fitted to the binned lags versus frequency distribution above 800~Hz yields -8.9 $\pm~2.3\mu$s\footnote{We note that in order to avoid using non-independent measurements we recomputed the time lags obtained from shifting-and-adding the lower kHz QPO without using the segments where both QPOs had already been detected.}. A trend for the hard lag to increase with frequency is also suggested. The same procedure can be applied to segments containing upper kHz QPOs. This then extends the frequency range for the lags of the lower kHz QPO towards the lowest frequencies. Those lags are also reported in Fig. \ref{lag_vs_freq}, showing that the lags measured, although with large error bars, may be consistent with zero, possibly breaking the trend for the soft lags to increase when the frequency decreases. Higher quality data are required before firm conclusions can be drawn.

\subsection{The energy spectrum of the lags}

\begin{figure}
\epsscale{0.65}
\plotone{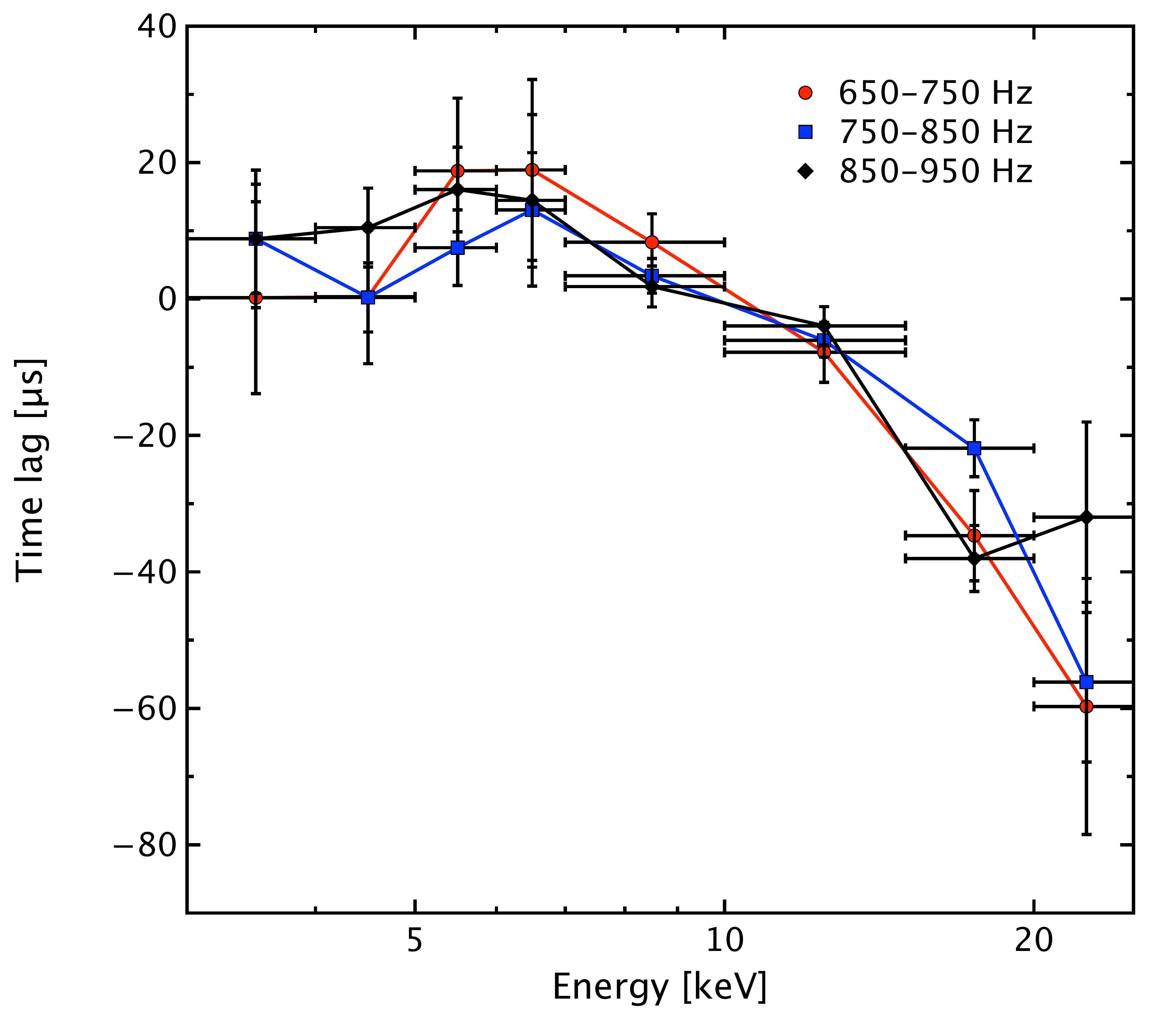}
\plotone{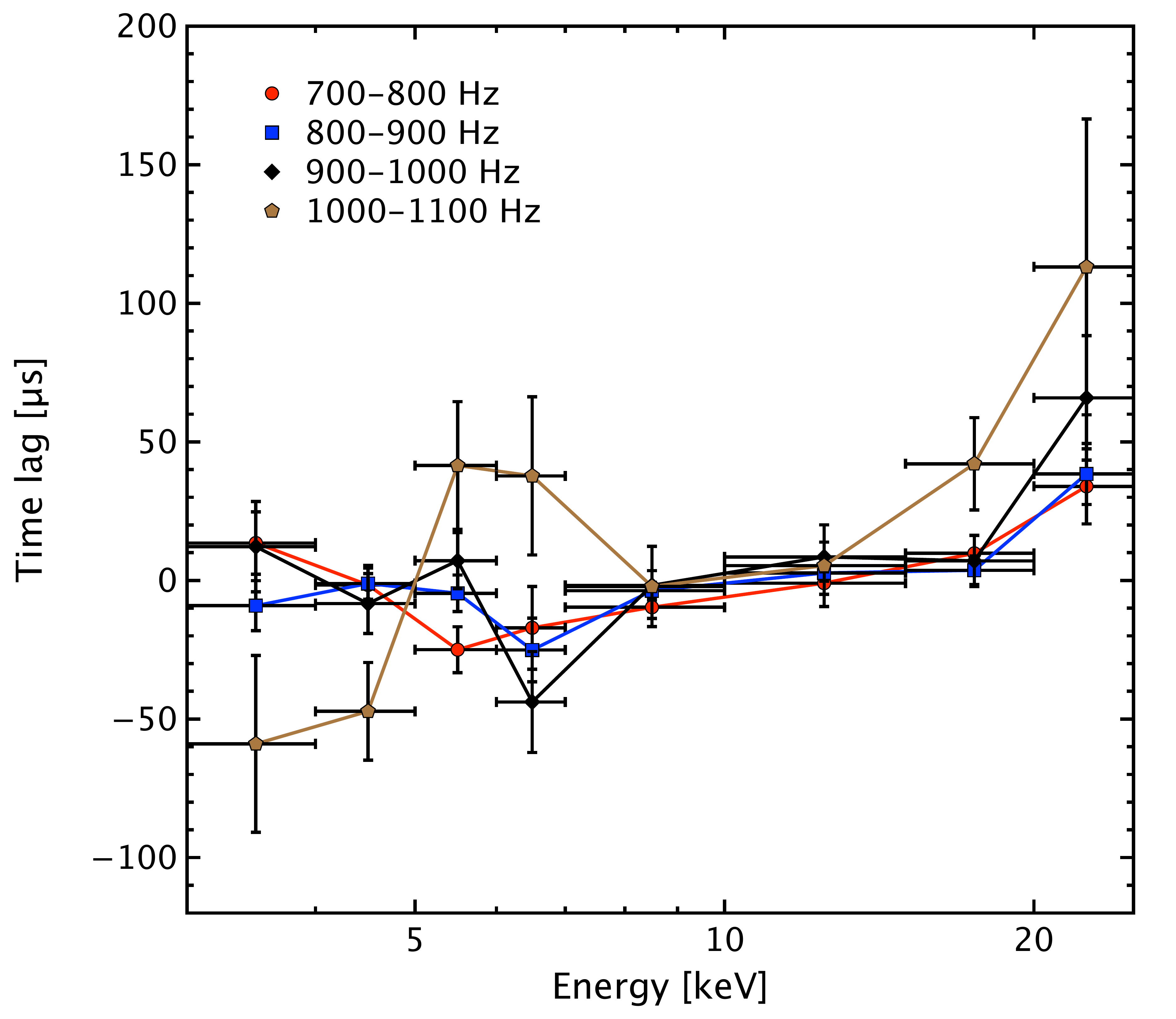}
\caption{Lag-energy spectra of the lower (top) and upper (bottom) kHz QPOs of \gx, obtained by averaging cross-spectra within different frequency intervals. We note that these spectra should be looked at in relative units: a zero lag value only means that the given energy band is in phase with the reference band and its position in the spectra depends on the choice of the latter.\label{lag_spectra}}
\end{figure}

\begin{figure}
\epsscale{1}
\plotone{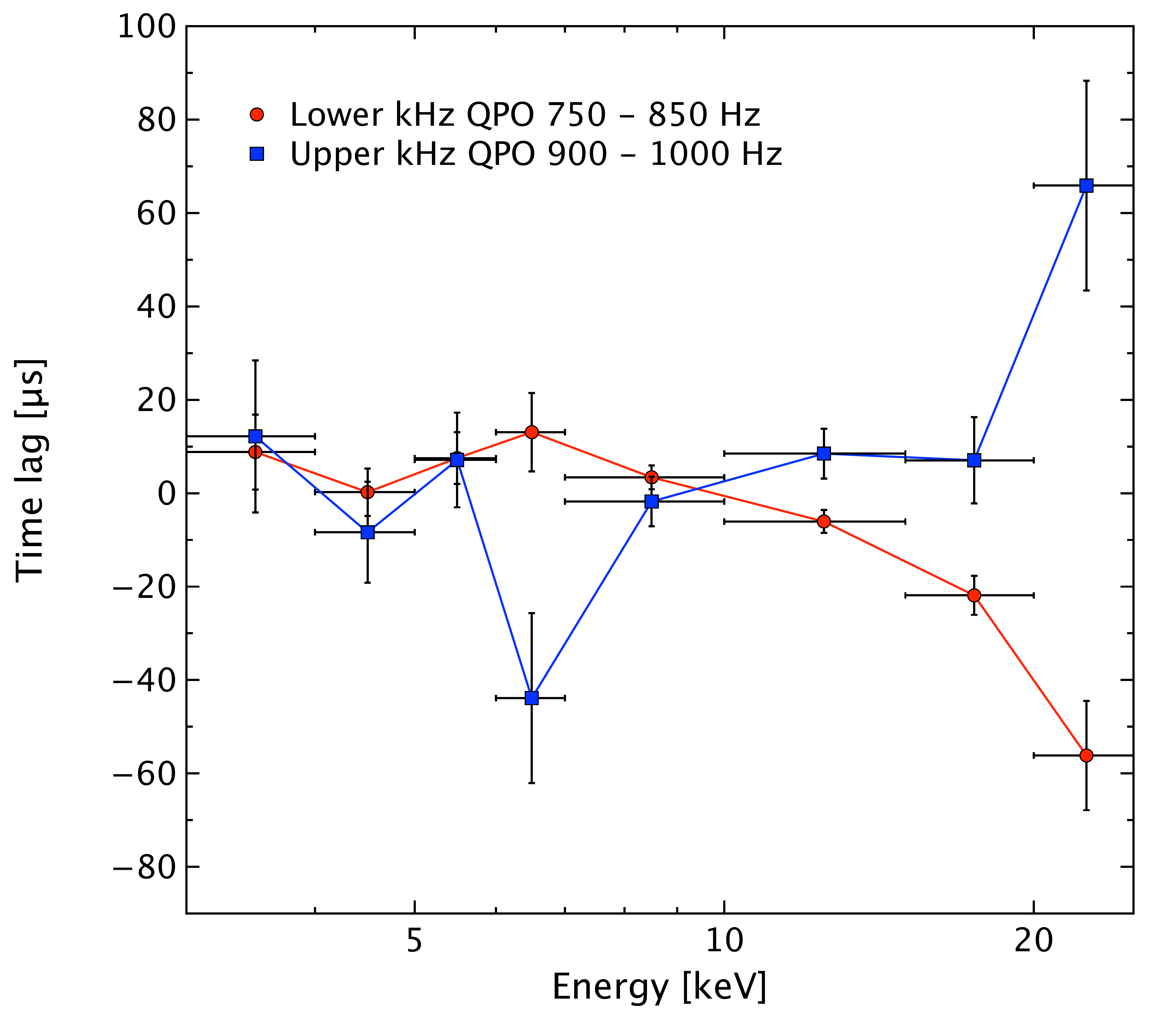}
\caption{Comparison of the lag-energy spectra obtained for the lower kHz QPO between 750 and 850~Hz (red circles), and for the upper kHz QPO between 900 and 1000~Hz (blue squares). We note that these spectra should be looked at in relative units: a zero lag value only means that the given energy band is in phase with the reference band and its position in the spectra depends on the choice of the latter.\label{comp_lag_spectra}}
\end{figure}

We now wish to examine how the lags vary with energy. In each energy bin, {using the same procedure as described above}, the  lag is computed between the light curve in that bin and the light curve in the reference energy band (3-25\footnote{We cut our analysis at 25~keV in order to avoid the background dominated part of the spectra.} keV), where the signal to noise ratio of the QPO detection is the highest. The light curve in the energy bin considered is subtracted from the reference light curve to ensure that Poisson noise remains uncorrelated \citep{Uttley:2011,Uttley:2014}. A mean lag-energy spectrum can be computed from the shifted-and-added cross spectrum over a given set of QPO segments. To improve the statistical errors on the lag-energy spectrum, we have used relatively large energy bins and grouped the data within adjacent frequency intervals. This way, minor changes in energy channel boundaries over the course of the observations considered in the average are small compared to the width of the energy bin. The lag-energy spectra of the lower and upper kHz QPOs of \gx\ are shown in Fig.~\ref{lag_spectra}.  It is striking that the overall shape of the lag-energy spectra of the two QPOs are very different (as illustrated by Fig.~\ref{comp_lag_spectra}). As for \4u\ and \1636\, the lag energy spectra of the lower kHz QPO of \gx\ show a smooth decrease with energy. At the lowest energy, the lag is positive, indicating that on average soft photons arrive after the broad continuum, while at the highest energy the lag is negative, indicating that hard X-ray photons on average reach the observer before the photons from the broad continuum. Although the energy resolution of the lag-energy spectrum is degraded as a result of the energy binning, there is no clear feature around 6-7 keV. Trying narrower binning degrades the statistics and prevents any features to be seen around these energies. As suggested by the flatness of the frequency dependency of the lower kHz QPO lags shown in Figure \ref{lag_vs_freq}, the lag-energy spectrum does not show much dependence on frequency: the three frequency intervals considered in the analysis yield three lag-energy spectra consistent with one another within error bars. As for the lag-energy spectrum of the upper kHz QPO, the shape is very different, in the sense that the lags show a plateau consistent with zero and then may rise with energy above $\sim 10$ keV. The same trend was present in the lag energy spectra of the upper kHz QPOs of \4u\ and \1636\, although with a larger scatter \citep{de-Avellar:2013}. One striking feature of Figure \ref{lag_spectra} is the last frequency interval (1000-1100 Hz) in which the lag-energy spectrum breaks the trend, and shows a clear bump between 5 and 10 keV. 

\subsection{Covariance spectra}

\begin{figure}
\epsscale{0.65}
\plotone{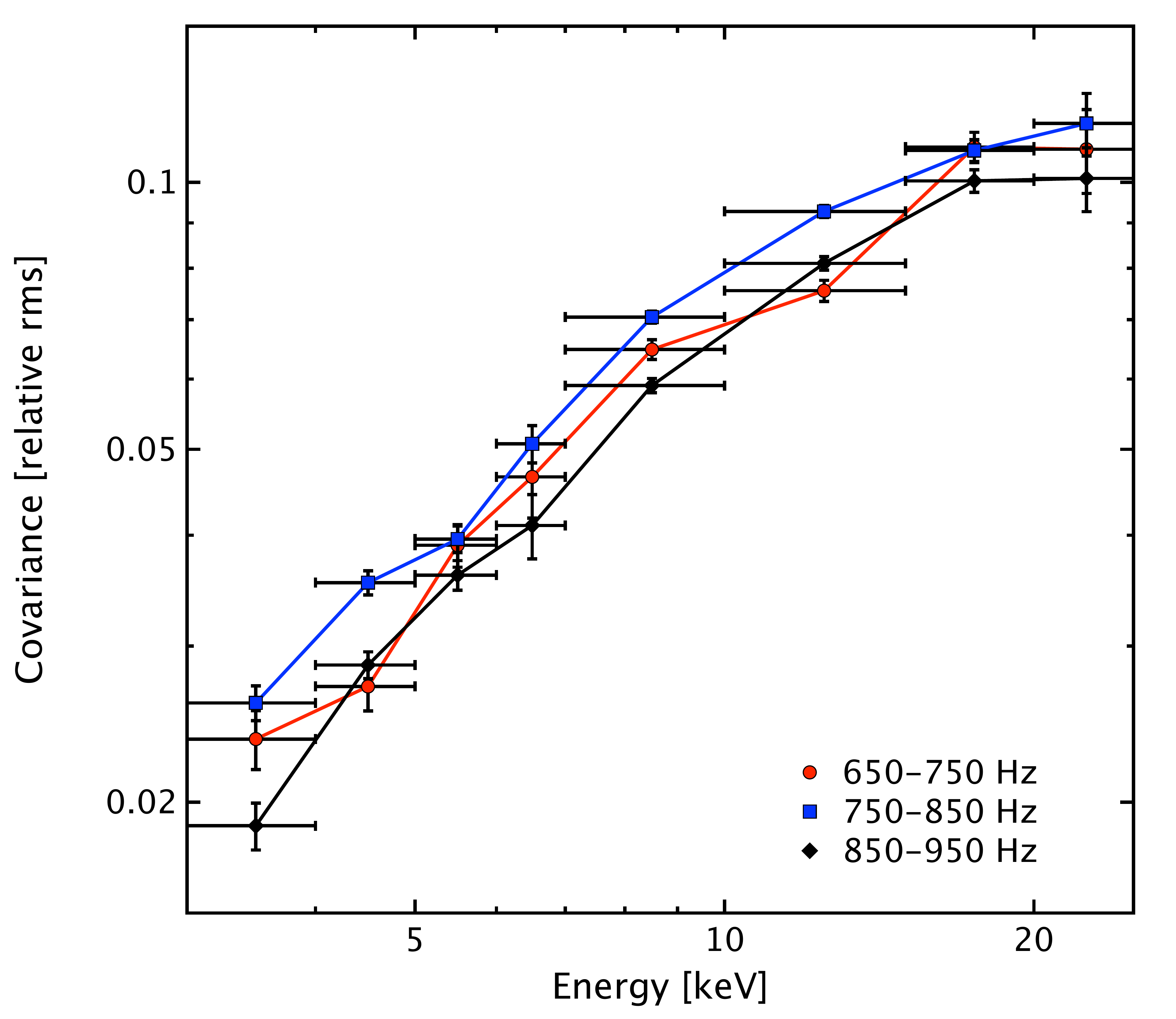}
\plotone{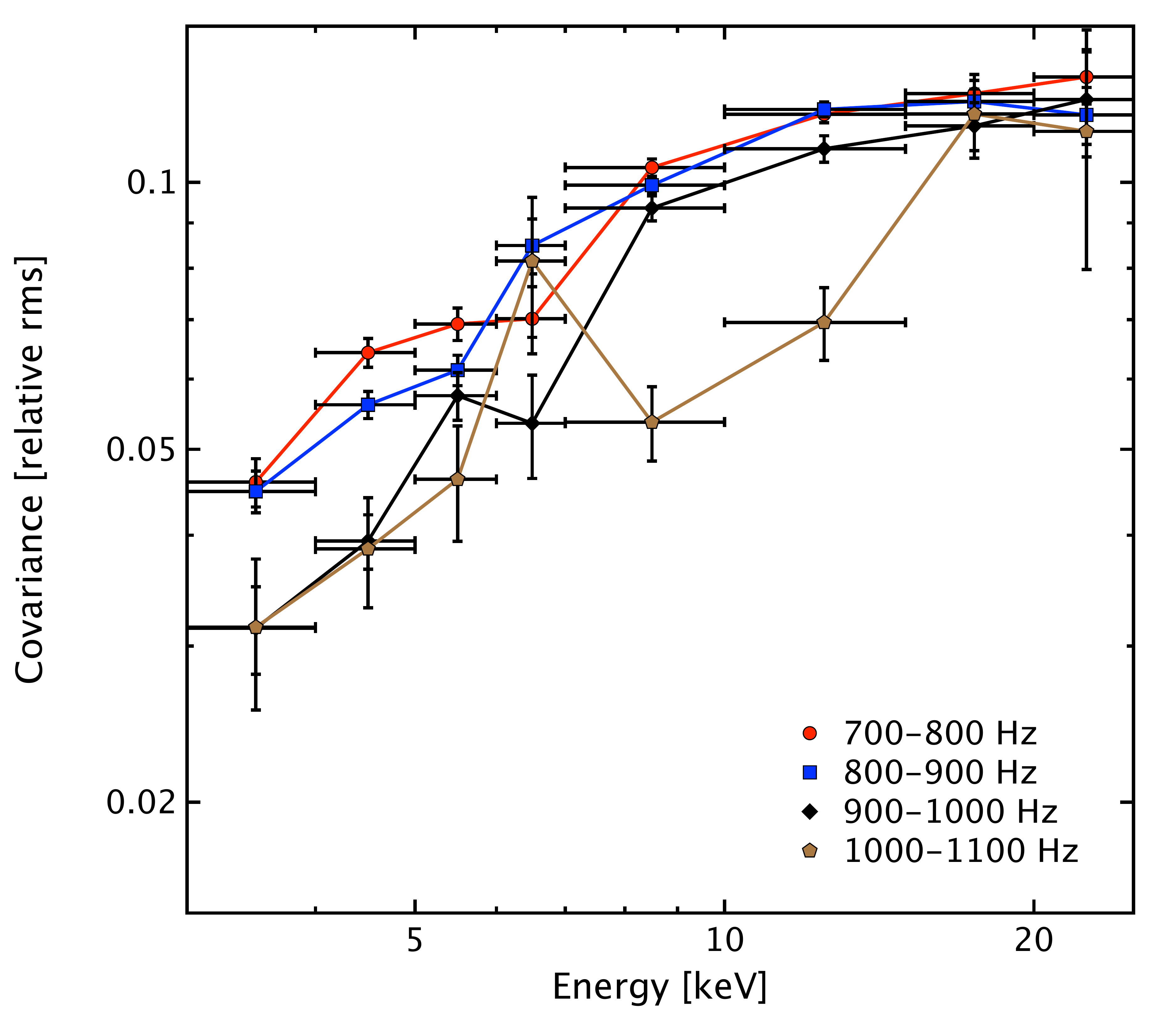}
\caption{Fractional (relative rms) covariance spectra of the lower (top) and upper (bottom) kHz QPOs of \gx, obtained by averaging cross-spectra within different frequency intervals. \label{cov_spectra}}
\end{figure}

\begin{figure}
\epsscale{1}
\plotone{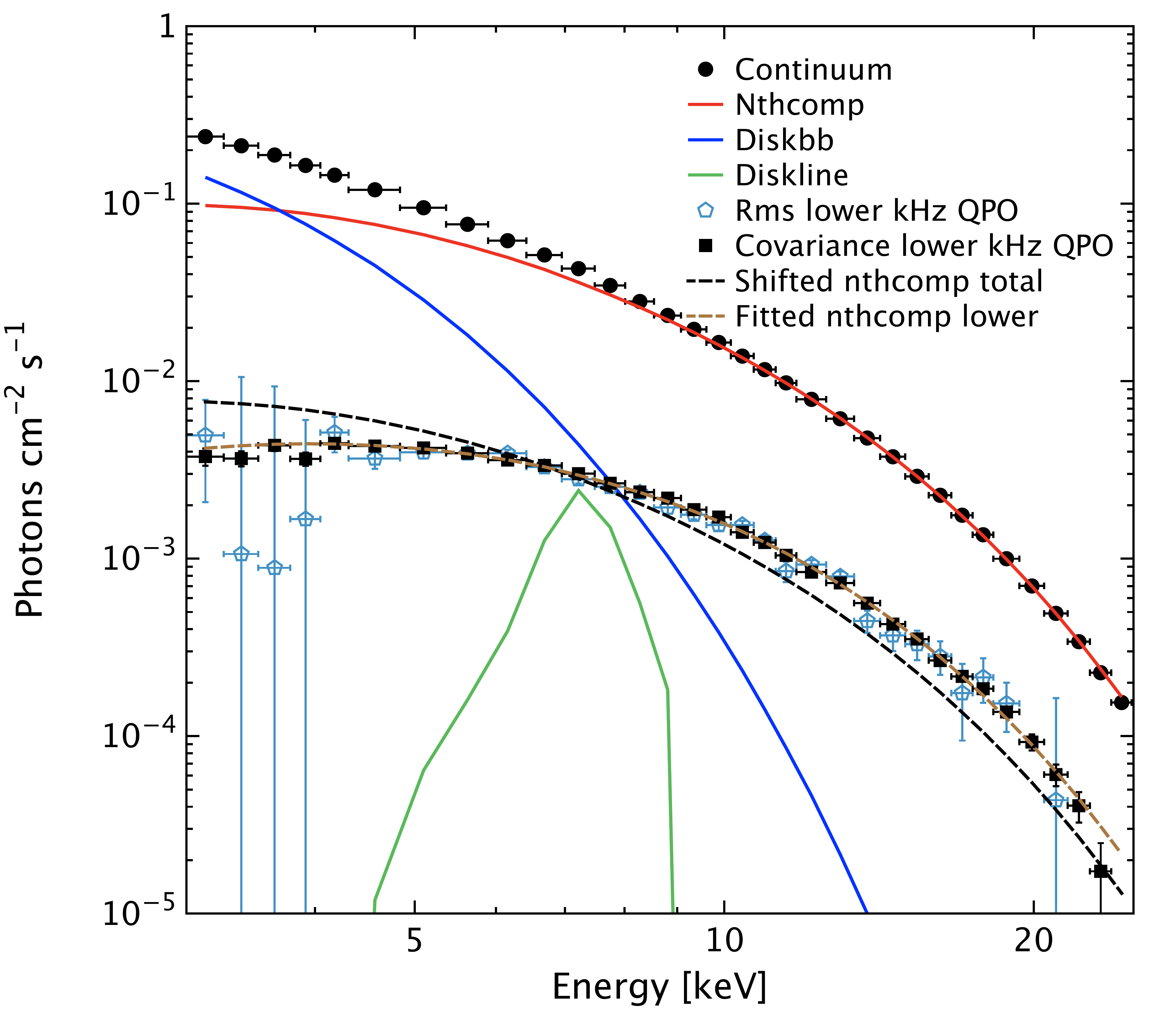}
\caption{Comparison of the decomposition of the continuum spectrum (black circles) obtained from the first ObsID of \4u\ (10072-05-01-00) with the covariance (black squares) and rms spectra (open points) of the lower kHz QPO detected in the same data segment. This decomposition was obtained from the joint fit of the covariance and continuum spectra ($\chi^2$ = 94.89 (56)). The continuum spectrum includes a Comptonization component (\texttt{nthcomp}, solid red line, $\Gamma = 1.96^{2.02}_{1.90}$, $kT_{e} = 2.61^{2.65}_{2.58}$ keV, $kT_{seed} = 1.10^{1.28}_{0.92}$ keV), a disk thermal emission (\texttt{diskbb}, solid blue line, $kT_{in} = 0.98^{1.07}_{0.88}$ keV), and a fluorescence iron line (\texttt{diskline}, solid green line, $E_{line} = 7.20^{7.31}_{7.10}$ keV). The dashed maroon line corresponds to the Comptonization component of the continuum spectrum fitted to the covariance spectrum allowing only the normalization and the seed photon temperature to vary. The seed photon temperature was thus found higher in the covariance spectrum than in the mean spectrum: $kT_{seed} = 1.84^{1.87}_{1.82}$ keV. The dashed black line corresponds to the same fit but allowing only the normalization to vary. This figure also illustrates the better statistics of the covariance spectrum with respect to the RMS spectrum.\label{compar_covar_rms_1608}}
\end{figure}

As a by-product of the above analysis, one can also compute the frequency resolved covariance spectra of the two kHz QPOs, in the same frequency intervals and over the same energy bins. The covariance spectrum measures the amplitude of the variability in each energy bin which is correlated with the variations in the reference band \citep[][for details]{Wilkinson:2009,Uttley:2011,Uttley:2014}. In the case where the variability is spectrally coherent, i.e. variations are strongly correlated across all energies (which has been reported for the lower kHz QPO by \citet{de-Avellar:2013}), the covariance spectrum is equivalent to the rms spectrum \citep[e.g.][]{Gilfanov:2003}, albeit with significantly better signal-to-noise (see Fig.~\ref{compar_covar_rms_1608}). The covariance spectra are shown in Fig.~\ref{cov_spectra}, in which they are normalized by the mean spectra for the same data sets, i.e. to relative RMS units. These are the first covariance spectra ever shown for kHz QPOs. 

\citet{Gilfanov:2003} found that the rms spectrum of the QPOs was consistent with the Comptonized blackbody emission assumed to originate from the boundary layer, with no obvious contribution from the accretion disk blackbody emission.  Our covariance spectra are consistent with this overall behavior, with a drop in fractional covariance towards lower energies, and a flattening at higher energies, which could be due to the presence of a constant disk component diluting the harder variability (see Fig. \ref{cov_spectra}). To further confirm this, we performed a spectral deconvolution of all the covariance spectra obtained from the different ObsIDs. After grouping them by QPO type and frequency\footnote{Only ObsIDs with at least two segments of 1024~s containing a QPO were kept in order to obtain covariance spectra with sufficiently high signal to noise ratio.} to account for possible spectral changes, we fitted them simultaneously with a Comptonized emission model (\texttt{wabs*nthcomp} in \texttt{xspec 12.8}), i.e. in each bin as many covariance spectra as ObsID were simultaneously fitted\footnote{Some ObsIDs during which the response changes were split such that each spectrum has its correct response matrix. This is however rather rare and we chose to keep the ObsID denomination in the rest of the paper to avoid confusion.}. The hydrogen column density was fixed to N$_H=2.6 \times 10^{22}$ cm$^{-2}$ \citep{DAi:2006} and only the normalization was allowed to vary between observations inside a frequency bin. We thus obtained good fits (reduced $\chi^2 \sim 1.0-1.3$). Interestingly enough, the models found for the lower kHz QPOs were all compatible within error bars, suggesting that the spectral shape of this emission is constant with frequency.

\begin{deluxetable}{cccccccc}
	\tabletypesize{\scriptsize}
	\tablecaption{$\chi^2$(dof) values obtained from the joint fits of the mean and covariance spectra corresponding to the data segments with kHz QPOs using different methods. \label{comparison}}
	\tablecolumns{8}
	\tablewidth{0pt}
	\tablehead{\colhead{QPO type} & \colhead{Frequency} & \colhead{all linked} & \colhead{$kT_{e}$ free} & \colhead{$\Gamma$ free} & \colhead{$kT_{seed}$ free} & \colhead{all free} & \colhead{\#ObsID}}
	\startdata
	Lower & 650--750 & 1603(431) & 1375(430) & 1133(430) & 1051(430) & 1044(428) & 7\\
	~ & 750--850 & 2974(857) & 2680(856) & 2279(856) & 2116(856) & 2099(854) & 12\\
	~ & 850--950 & 1334(447) & 1089(446) & 770(446) & 701(446) & 692(444) & 7\\
 	\hline
	Upper & 700--800 & 4025(915) & 4025(914) & 3991(914) & 3933(914) & 3920(912) & 14\\
	~ & 800--900 & 3612(881) & 3554(880) & 3355(880) & 3190(880) & 3173(878) & 13\\
	~ & 900--1000 & 1232(317) & 1179(316) & 1070(316) & 1011(316) & 998(314) & 4\\
	~ & 1000--1100 & 194(95) & 177(94) & 178(94) & 185(94) & 172(92) & 2\\
	\enddata
	\tablecomments{The joint fits were conducted using the \texttt{xspec} model \texttt{wabs*(ntchomp+diskline+diskbb)}. For the covariance spectra, only the normalization of the \texttt{nthcomp} component was kept non-null. Depending on the cases, either the electron temperature ($kT_{e}$), the $\Gamma$ shape parameter, the seed photon temperature ($kT_{seed}$), none, or all the parameters were allowed to differ between the covariance and the mean spectra. In all cases, the hydrogen column density was fixed to N$_H=2.6 \times 10^{22}$ cm$^{-2}$ \citep{DAi:2006}. The last column indicates how many ObsIDs were combined in each frequency bin.}
\end{deluxetable}

\begin{deluxetable}{ccccccccc}
	\tabletypesize{\scriptsize}
	\tablecaption{Results of the joint fit of the mean and covariance spectra corresponding to the data segments with kHz QPOs.	\label{overall}}
	\tablecolumns{9}
	\tablewidth{0pt}
	\tablehead{\colhead{QPO type} & \colhead{Frequency} & \colhead{$kT_{seed}$(mean)} & \colhead{$kT_{seed}$(cov)} & \colhead{$\Gamma$} & \colhead{$kT_e$} & \colhead{$E_{line}$} & \colhead{$kT_{in}$} & \colhead{$\chi^2$ (dof)}\\
	\colhead{} & \colhead{(Hz)} & \colhead{(keV)} & \colhead{(keV)} & \colhead{} & \colhead{(keV)} & \colhead{(keV)} & \colhead{(keV)} & \colhead{}}
	\startdata
	Lower & 650--750 & 1.25$_{1.21}^{1.29}$ & 2.17$_{2.14}^{2.21}$ & 2.20$_{2.17}^{2.23}$ & 3.51$_{3.49}^{3.55}$ & 6.58$_{6.55}^{6.60}$ & 0.83$_{0.80}^{0.85}$ & 1051 (430)\\
	~ & 750--850 & 1.12$_{1.09}^{1.16}$ & 2.08$_{2.06}^{2.11}$ & 2.12$_{2.09}^{2.14}$ & 3.28$_{3.26}^{3.31}$ & 6.56$_{6.55}^{6.58}$ & 0.73$_{0.70}^{0.76}$ & 2116 (856)\\
	~ & 850--950 & 1.12$_{1.07}^{1.16}$ & 2.17$_{2.14}^{2.20}$ & 2.01$_{1.99}^{2.04}$ & 3.07$_{3.05}^{3.09}$ & 6.65$_{6.62}^{6.67}$ & 0.74$_{0.71}^{0.77}$ & 701 (446)\\
 	\hline
	Upper & 700--800 & 1.43$_{1.41}^{1.46}$ & 1.80$_{1.77}^{1.82}$ & 2.37$_{2.35}^{2.39}$ & 9.56$_{9.15}^{9.96}$ & 6.60$_{6.59}^{6.61}$ & 1.05$_{1.03}^{1.06}$ & 3933 (914)\\
	~ & 800--900 & 1.28$_{1.25}^{1.30}$ & 1.84$_{1.82}^{1.87}$ & 2.35$_{2.33}^{2.36}$ & 6.00$_{5.88}^{6.09}$ & 6.58$_{6.56}^{6.59}$ & 0.87$_{0.85}^{0.89}$ & 3190 (880)\\
	~ & 900--1000 & 1.14$_{1.11}^{1.16}$ & 1.88$_{1.83}^{1.93}$ & 2.32$_{2.30}^{2.34}$ & 4.56$_{4.49}^{4.66}$ & 6.53$_{6.51}^{6.56}$ & 0.68$_{0.66}^{0.70}$ & 1011 (316)\\
	~ & 1000--1100 & 1.25$_{1.21}^{1.33}$ & 2.15$_{1.84}^{2.47}$ & 2.21$_{2.13}^{2.30}$ & 3.57$_{3.52}^{3.62}$ & 6.66$_{6.59}^{6.71}$ & 0.80$_{0.74}^{0.83}$ & 185 (94)\\
	\enddata
	\tablecomments{The joint fits were conducted using the \texttt{xspec} model \texttt{wabs*(ntchomp+diskline+diskbb)}. For the covariance spectra, only the normalization of the \texttt{nthcomp} component was kept non-null. Only the normalization and the seed photon temperature were allowed to differ from mean spectra. $kT_{seed}$ corresponds to the temperature of the seed photons of the \texttt{nthcomp} component, $\Gamma$ to a shape parameter of this component, $kT_e$ to the temperature of the Comptonizing electrons, $E_{line}$ to the centroid energy of the iron line (only used during the fit of the mean spectrum), and $kT_{in}$ to the inner temperature of the accretion disk (only used during the fit of the mean spectrum). The sub- and super-scripts give the 1$\sigma$ error {\it ranges} (i.e. best-fitting value plus lower and upper error bars respectively).}
\end{deluxetable}

\begin{figure}
\epsscale{1}
\plotone{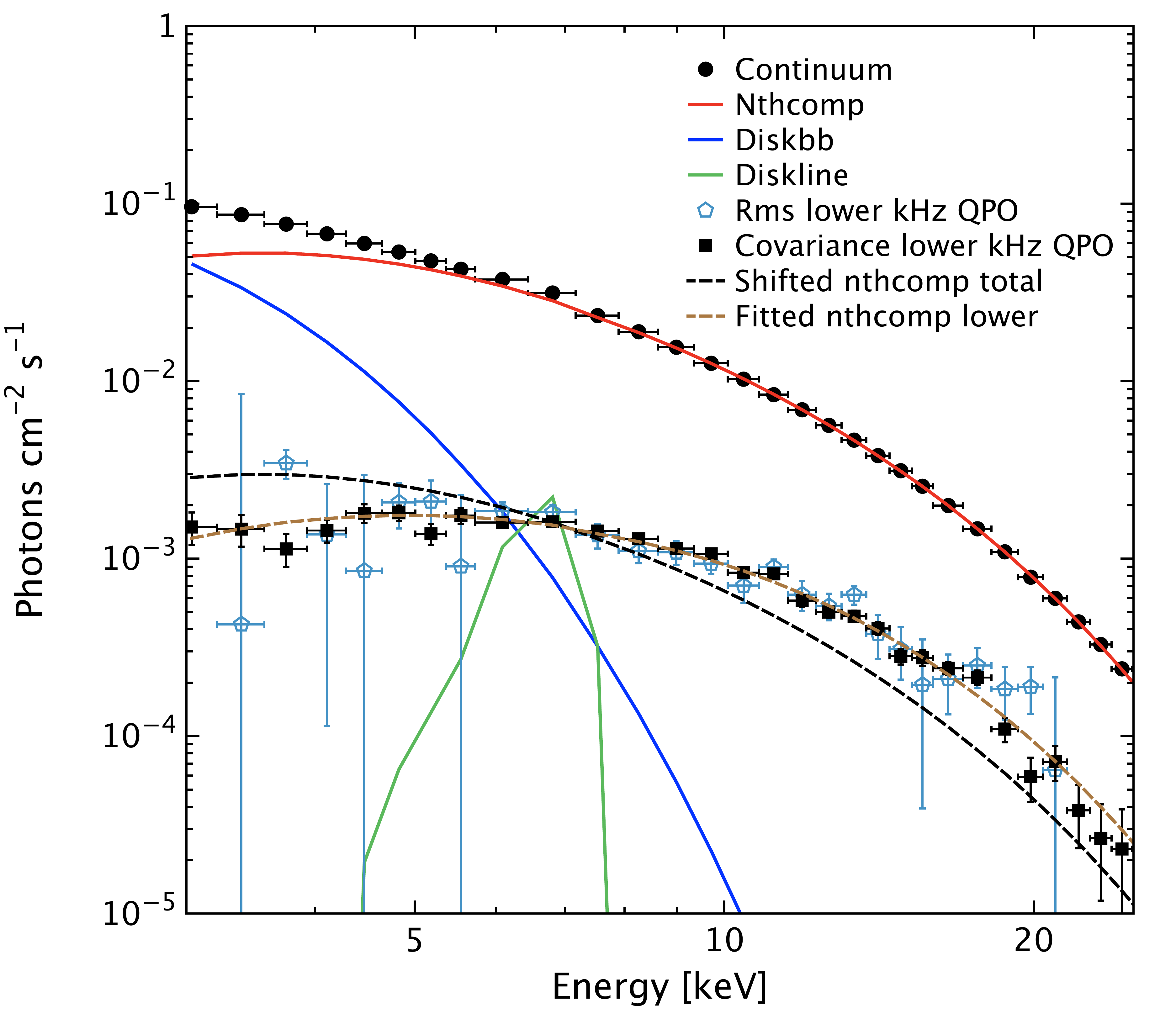}
\caption{Comparison of the decomposition of the continuum spectrum (black circles) obtained from one ObsID of \gx\ (20083-01-03-020) with the covariance (black squares) and rms spectra (open points) of the lower kHz QPO detected in the same data segment. This decomposition was obtained from the joint fit of the covariance and continuum spectra ($\chi^2$ = 56.69 (52)). The continuum spectrum includes a Comptonization component (\texttt{nthcomp}, solid red line, $\Gamma = 2.09^{2.16}_{2.01}$, $kT_{e} = 3.13^{3.19}_{3.05}$ keV, $kT_{seed} = 1.20^{1.29}_{1.08}$ keV), a disk thermal emission (\texttt{diskbb}, solid blue line, $kT_{in} = 0.75^{0.80}_{0.68}$ keV), and a fluorescence iron line (\texttt{diskline}, solid green line, $E_{line} = 6.60^{6.67}_{6.52}$ keV). The dashed maroon line corresponds to the Comptonization component of the continuum spectrum fitted to the covariance spectrum allowing only the normalization and the seed photon temperature to vary. The seed photon temperature was thus found higher than in the mean spectrum: $kT_{seed} = 2.24^{2.30}_{2.19}$ keV. The dashed black line corresponds to the same fit but allowing only the normalization to vary.\label{compar_covar_rms_1728}}
\end{figure}

We now wish to see whether the spectral shape of the covariance spectra can be linked to a component in the total emission spectrum of the source (i.e. the mean, time-averaged, spectrum), which may also contain components which do not vary on the kHz QPO time-scale (e.g. \citealt{Gilfanov:2003}). The covariance spectra are therefore jointly fitted with their corresponding mean spectra. We use the common decomposition of the total emission into an accretion disk, a Comptonized component, and a relativistically broadened iron line (in \texttt{xspec}: \texttt{wabs*(diskbb+nthcomp+diskline)}) for the mean spectra \citep{Barret:2013}. Following the same idea as \cite{Gilfanov:2003} who tried to identify the shape of the rms spectrum of the lower kHz QPO with the disk-subtracted spectrum of the source, only the Comptonized component is retained for the covariance spectra. The hydrogen column density is fixed, as well as the parameters of the iron line, using an inclination angle of $i=50$~deg \citep{Shaposhnikov:2003} and the default model parameters set otherwise (a disk emissivity dependence in $R^{-2}$, $R_{in} = 10~r_g$, and $R_{out} = 1000~r_g$). During the fits, the normalization of each component is left free between ObsIDs. We note that letting only the global normalization vary does not change the fitted parameters significantly, but only degrades the quality of the fits. Forcing all the parameters of the Comptonization component to be the same for the mean and covariance spectra yields bad fits, but freeing the seed photon temperature corrects them: once this is done, the fits are as good as those obtained with all the parameters free, i.e. untying other parameters from the mean spectrum does not significantly change the fit statistic (see Table \ref{comparison}). Note that the seed photon temperature is found to be systematically higher for the covariance spectra than for the mean spectra (see Table \ref{overall}, and Fig.~\ref{compar_covar_rms_1728} for an illustrative example on one ObsID). 

The chosen decomposition for the continuum emission is one among the many others that work well with neutron star binaries \citep{Lin:2007}. To test the robustness of our results against the choice of the continuum, we replaced the disk emission by a simple black body (\texttt{bbodyrad}) or the Comptonization component by a broken power law (\texttt{bknpower}). Very similar results were found after the replacement of the thermal component: allowing a higher seed photon temperature in the covariance spectra than in the continuum corrects the fits and yields similar final $\chi^2$ values as with \texttt{diskbb} (albeit slighly larger in 6 out of 7 cases). However, when the Comptonization component is replaced by a broken power law, fits of comparable quality as before are only obtained when the continuum and covariance spectra are completely untied. We also note that the fits with the parameters of the covariance tied to the continuum yield systematically worse $\chi^2$ values in both cases. Our results were therefore found robust against the change of the thermal component, but in the absence of a significant change in all the final $\chi^2$ values of the fits, it is difficult to use this analysis as a way to disentangle between the various spectral decompositions that can fit neutron star spectra.

It is important to mention that these fits using a single set of model parameters were made while combining the emission of the source from many observations grouped together using relatively large kHz QPO frequency ranges and therefore slightly different source spectra \citep{Di-Salvo:2001}. This explains the relatively poor $\chi^2$ values obtained here\footnote{We also note that the exposure times in the spectra were not deadtime corrected and that this correction would lead to a reduction in the $\chi^2$ values.}. Consistent with this explanation, we fitted the mean and covariance spectra by pairs, and found total reduced $\chi^{2}$ for each frequency bin typically around 1 for the different frequency/QPO combinations, when the seed photon temperature of the two spectra were untied. As before, the seed photon temperature was systematically found higher in the covariance spectra than in the mean spectrum, with values consistent with the previous analysis (simultaneous fits of all the spectra in one frequency bin).

In order to verify if the feature seen in \gx\ is unique to this source or shared with others, we have carried out the same analysis on the first ObsID of \4u, namely 10072-05-01-00, which is known to contain QPO data of particularly high quality \citep[see e.g.][]{Barret:2005}. A similar result was thus found (see Fig.~\ref{compar_covar_rms_1608}), with a decomposition of the continuum consistent within error bars with the one shown in \citet{Barret:2013}. This supports the idea that this may be a common feature of the kHz QPOs and that their spectra are systematically harder than the Comptonization component of the continuum.

\section{Discussion}

The main observational results of our analysis can be summarized as follows:
\begin{itemize}
	\item The lag-energy spectra of the lower and upper kHz QPOs of \gx\ are systematically different: the lower kHz QPO shows soft variations lagging hard variations, as seen in other sources, while the upper kHz QPO shows either a flat lag-energy spectrum or hard variations lagging softer variations.
	\item The shapes of the lag-energy spectra depend only weakly on the QPO frequency.
	\item The covariance spectra of both kHz QPOs are harder than the continuum Comptonization component fitted to the mean spectra of the persistent emission. Allowing the seed photon temperature to be different in the kHz QPO covariance spectra than in the mean spectra improves the fit significantly, with a higher seed temperature seen in the kHz QPO spectra.
\end{itemize}

The simplest interpretation for the lag-energy spectra to be different is that despite their overall spectral similarity, the two QPOs are generated by different mechanisms.  We now consider a number of possible mechanisms to explain the lags and spectral variability and comment on their viability in the light of our results.

As noted by \citet{Barret:2013}, the soft lags seen in the lower kHz QPO might be produced by the light-travel delay associated with thermal reverberation (i.e. X-ray heating) of the accretion disk by the boundary-layer continuum. However, this cannot be the whole story: the lag-energy spectrum continues to drop, well above energies where the disk blackbody emission should contribute, suggesting that at least some component of the lags is intrinsic to the primary Comptonized continuum which dominates at these energies.  For the upper kHz QPO, reverberation of the variable boundary layer emission off the disk to produce a lagging reflection component, could be a viable mechanism to explain the lags, given the relative flatness, and then hardening of the lag-energy spectra. The observed lags are consistent with those expected from a disk extending close to the surface of a neutron star: the lags of order 100~$\mu$s correspond to light-travel distances of 30~km.  Note that one would have to correct for the primary continuum source geometry as well as dilution of the lag by the variable primary continuum to arrive at a reliable estimate for the disk inner radius (e.g. see \citealt{Uttley:2014}).

Next, we consider variations linked to thermal Comptonization in the boundary layer, which seems to explain the energy-spectral shape of the QPOs well (this paper and \citealt{Gilfanov:2003}). Recently, \citet{Kumar:2014} considered different mechanisms linked to thermal Comptonization to cause energy-dependent time-delays in kHz QPOs, including predictions for the rms-spectrum (equivalent in this case to the covariance) as well as the lags, which we can compare with our results.  Firstly, the case of pure seed photon temperature oscillations seems to be ruled out, since it predicts a pivoting in the spectrum leading to a deep minimum in the fractional rms at the pivot energy, between 10-20 keV, which is not seen in our data.  In fact, we see no evidence for significant dips in fractional rms at energies where the Comptonized component dominates the mean and covariance spectra (see Fig.~\ref{cov_spectra}). The covariance and mean spectra at these energies (above where the disk contributes to the mean) are well fitted by the same-spectral model, with the only difference being the seed temperature. The combination of different seed-photon temperatures with an additional constant disk component contributing to the mean spectrum can explain the observed energy-dependence in fractional rms units of the covariance.

\citet{Kumar:2014} also investigated the case of oscillations in the heating rate of the Comptonizing region (see also \citealt{Lee:2001}).  Interestingly, such oscillations can produce either hard or soft lags, with soft lags being produced if a significant fraction of hard photons can impinge on and heat the seed-photon source.  In this latter case, the oscillating region must also be rather compact ($\sim$1~km) in order to produce significant variability at the QPO frequencies \citep{Kumar:2014}.  The existence of a compact oscillating region might be supported by our finding of a higher seed-photon temperature for the covariance spectra of both lower and upper kHz QPOs than for the mean spectrum, which suggests that the oscillating component of the Comptonizing region sees a hotter, more compact source of seed-photons than the non-oscillating part.  However, oscillations in coronal heating rate also predict a fractional rms which continually increases with energy above 10~keV, whereas the data show that the fractional rms flattens above 10~keV.  \citet{Kumar:2014} note similar rms-spectral behaviour in \4u\ and suggest that an additional hard constant component may flatten the fractional rms at high energies.  We find no evidence for such a component in our spectral fits: the mean spectrum, which will contain any constant components (such as the disk emission) can be simply explained by the same spectral model that produces the QPO covariance spectra, with the only difference being the seed-photon temperature and the presence of the disk at lower energies.  No additional hard component is required to fit the mean spectrum.

None of the simple models we have considered satisfactorily explains all of the spectral-timing features of both lower and upper kHz QPOs.  Therefore it may indeed be more likely that although the upper and lower kHz QPOs share some broad similarities (e.g. a thermal-Comptonized spectrum), the mechanisms producing each are fundamentally different in some way.  For example, if the lag-energy spectrum in the upper kHz QPO is dominated by the reverberation signal, this would imply that there is little intrinsic spectral variation and corresponding lags from the primary continuum, with only the continuum normalisation changing. This might then suggest that the upper kHz QPO corresponds to a simple variation in luminosity, perhaps driven by fluctuations in the accretion rate on to the neutron star boundary layer, generated in the innermost radii of the accretion disc.  There is already evidence to suggest that accretion flow variability drives the broader noise components seen at lower-frequency (e.g. \citealt{Uttley:2004,Uttley:2011}), and it seems more probable that the lower-coherence upper kHz QPO has such an origin than the much more coherent lower kHz QPO, which would be difficult to produce in a turbulent and shearing accretion flow \citep{Barret:2005}.  

In fact, recent work by \citet{Bult:2015} strongly links the upper kHz QPO to an oscillation in the inner disk, by showing that the accretion-powered pulsations of the millisecond X-ray pulsar SAX~J1808.4-3658 are suppressed when the upper kHz QPO frequency exceeds the spin-frequency (suggesting a centrifugal boundary).  The signal is ultimately generated in the inner disk, which generates only a very small fraction of the observed disk emission, so the disk emission remains constant. The signal is thus only observed in the response of the boundary layer to the corresponding accretion variations.

In contrast, the lower kHz QPO might then be associated with a more coherent signal produced in a compact region of the boundary layer itself and thus able to produce relatively coherent variations due to the small scale-size. For instance internal oscillations in the heating rate of the boundary layer could produce the observed soft lags as envisaged by \citet{Lee:2001} and \citet{Kumar:2014}, although clearly more work is needed to explain the shape of the rms spectrum. However, even if, as seems likely, the lower and upper kHz QPOs are generated by different mechanisms, given their related frequencies, it is also likely that both QPOs track some underlying property of the flow, perhaps linked to the global accretion rate.

\section{Conclusions}

In this paper, we presented the first exhaustive cross-spectral analysis of both kHz QPOs in \gx, which showed that the lower and upper kHz QPOs have distinct lag properties. Even if this could be an indication that the emission mechanisms and/or origin of each QPO are fundamentally different, a careful modeling of the light travel in neutron star low-mass X-ray binaries would be needed to interpret unambiguously the detailed shape of the lag spectra we measured. From the computation of high signal to noise covariance spectra of both kHz QPOs, we also found further evidence that the emission of both QPOs is compatible with that of a Comptonized component \citep{Gilfanov:2003}. The fact that these spectra are better fitted with higher seed-photon temperatures than the same component in the mean spectra may also suggest that a particular, compact region of the boundary layer is responsible for the X-ray emission at the two QPO frequencies.

Our results further illustrate the power of joint timing and spectral analysis of low-mass X-ray binaries, and we believe this type of measurements should be extended to other sources in which kHz QPOs have been cleanly detected with the \emph{RXTE PCA}. This may bring further information on the origin of the oscillators producing the kHz QPOs, and on the mechanisms by which the X-rays are ultimately modulated, which is a necessary step before we can use these signals to probe strong field General Relativity. In this perspective, the need for better data is clear and a next generation dedicated timing mission like \emph{LOFT} \citep{Feroci:2012} would be decisive.

\begin{acknowledgements} 

This research has made use of data obtained through the High Energy Astrophysics Science Archive Research Center On-line Service, provided by the NASA/Goddard Space Flight Center. We are grateful to the referee for very useful comments that helped to strengthen the claims made in this paper.

\end{acknowledgements} 

\bibliography{Biblio}

\end{document}